\documentclass{elsart1p}

\usepackage{graphicx}
\usepackage{amssymb}

\usepackage{amsmath}
\usepackage{hyperref}

\begin{document}
\begin{frontmatter}
% use the corauthref command within \author for corresponding author
% footnotes;
\title{On Anisotropy of Ultra-High Energy Cosmic-Rays}
\author{Tamar Kashti}
\address{Weizmann Institute of Science, Israel}
%
%\ead{tamar.kashti@weizmann.ac.il}

\begin{abstract}
We briefly summarize our study on anisotropy of Ultra-High Energy Cosmic-Rays (UHECRs), in which we
define a statistics that measures the correlation between UHECRs and Large Scale Structure (LSS).
We also comment here on recently published paper by Koers and Tinyakov that compared our statistics
to improved KS statistics.
\end{abstract}
\end{frontmatter}

\begin{figure}[tb]
\hbox{
\includegraphics[width=0.45\textwidth]{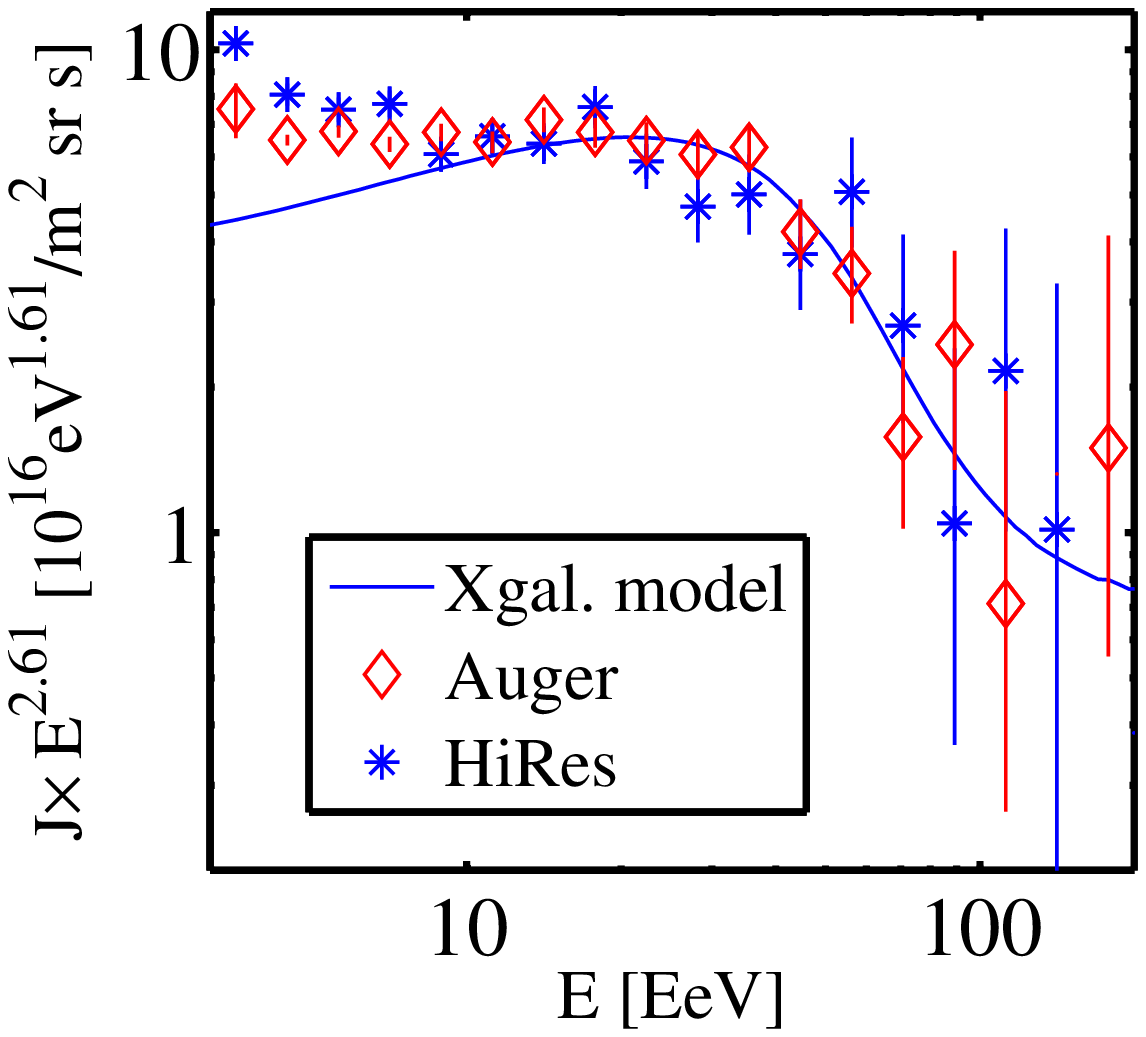}
\includegraphics[width=0.55\textwidth]{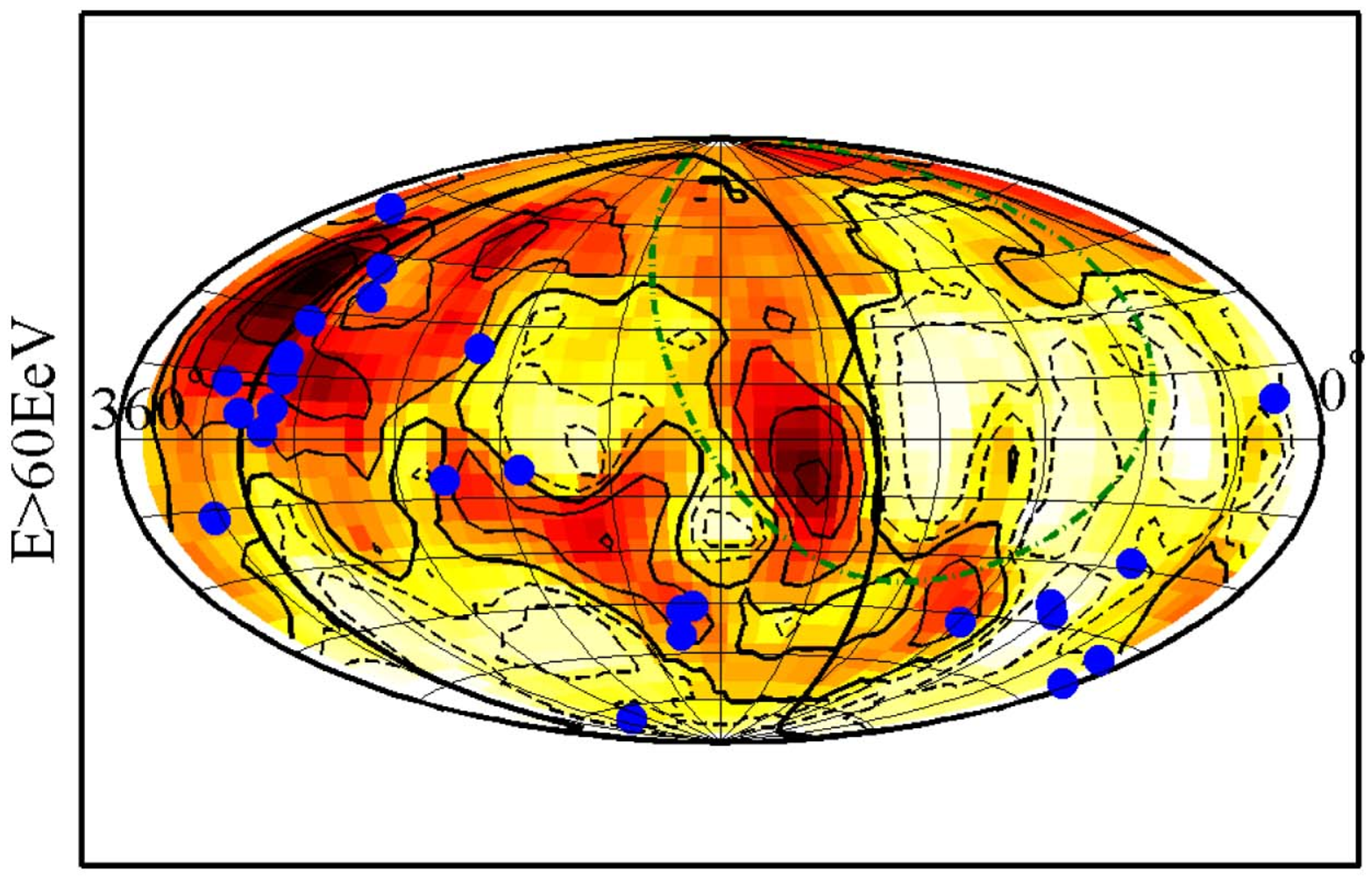}}
\caption{(a) The spectrum of HIRES and Auger after a shift of $\Delta E/E=+23\%$ in the calibration
of the absolute energy scale of the Auger experiment. The solid line is the spectrum that would be
generated by a cosmological distribution of sources of protons, with intrinsic spectrum $d\log
n/d\log E=-2$. %and redshift evolution following the star-formation rate, $\bar{s}(z)\propto
%(1+z)^3$.
(b) The positions of the 27 Auger events with energy exceeding $5.7\times10^{19}$~eV,
overlaid on the intensity map obtained in the biased model, in galactic coordinates.}
    \label{fig:Spec+Auger}
\end{figure}

%\section{Introduction}
% Introduction: puzzle, acceleration- sources (L), protons, GZK. - one page
The origin of cosmic rays %the Ultra-High Energy Cosmic Rays (UHECRs),
of energies $>10^{19}$~eV is a puzzle \cite{Bhattacharjee:1998qc,Nagano:2000ve,Waxman_CR_rev}. The
arrival direction of UHECRs show no correlation with the galactic disk, which point towards
extra-galactic origan. A suppression of the spectrum at $\sim 5\times 10^{19}$~eV is expected due
to %inelastic interactions of UHECRs with the cosmic microwave photons, the so called
the GZK suppression \cite{Greisen:1966jv,Zatsepin:1966jv}. The suppression was observed by HIRES
\cite{Abbasi:2007sv} and the new Auger Observatory \cite{Abraham:2008ru}, see
fig.~\ref{fig:Spec+Auger}(a). Thus, cosmic-rays with energies above $\sim 5\times 10^{19}$~eV can
reach us only from sources with distance below $\sim100$ Mpc. In these distances, the Universe is
not isotropic, implying that correlation with LSS can give us information on the origin of UHECRs.

There are few leading candidates as the sources of UHECRs. Assuming that UHECRs are protons, the
large magnetic fields needed for acceleration requires source luminosity of $L\gtrsim10^{12}
L_{\odot}\Gamma^2/\beta$ for protons with $10^{20}$~eV, where $L_\odot$ is the Sun luminosity and
$\Gamma$ is Lorentz factor of the magnetized plasma \cite{Waxman_CR_rev}.
%Assuming that the acceleration is due to magnetic fields in astrophysical sources, and that the
%accelerated particles are protons with energy $E_p$, one can derive a lower limit on the luminosity
%of the the accelerating source:
%\begin{align}
%L>10^{12} L_{\bigodot} \left(\frac{E_p}{10^20 \mbox{~eV}}\right)^2 \frac{\Gamma^2}{v/c}
%\end{align}
%where $L_{\bigodot}$ is the Sun luminosity, $v$ is the velocity of the magnetized plasma, and
%$\Gamma$ is its Lorentz factor \cite{Waxman_CR_rev}.
Therefore, the only known astrophysical sources that reaches these energies are Active Galactic
Nuclei (AGNs) and sources of Gamma-ray bursts. The anisotropy of UHECRs from these sources should
be correlated with LSS. Another possible source of UHECRs is the decay of new heavy particles
coming from top-down models \cite{Bhattacharjee:1998qc}, which predict isotropic signal.

% Model, statistics, results, Auger. - one page
%\section{Results}
In \cite{Kashti:2008bw}, we derive the expected all sky angular distribution of the UHECR
intensity, and defined a statistics $X_{C,UB}$ that measures the correlation between the predicted
and observed UHECR arrival direction distribution. Following \cite{Waxman:1996hp}, we consider a
model where the UHECR flux is produced by cosmological sources of protons tracing the large scale
galaxy distribution. We assume that the sources are intrinsically identical and that the number
density of sources is drawn from a Poisson distribution with an average given by
$b[\delta]\bar{s}(z)$, where $\bar{s}(z)$ is the average comoving number density of sources at
redshift $z$ and $b$ is some bias functional of the local fractional galaxy over density,
$\delta\equiv\delta\rho/\bar\rho$. The LSS galaxy density field is derived from the PSCz catalogue
\cite{Saunders:2000af}. For the bias functional we consider three models: an isotropic (I) model,
 $b[\delta]=1$; an unbiased (UB) model where the source
distribution traces the galaxy distribution with $b[\delta]=1+\delta$; and a biased (B) model,
% in which the UHECR source distribution is biased compared to the galaxy distribution with a threshold
%bias,
$b[\delta]=1+\delta$ for $\delta>0$ and $b[\delta]=0$ otherwise.

The statistics we defined in \cite{Kashti:2008bw}, which measures the correlation between predicted
and observed UHECR arrival direction distributions, is:
\begin{align*}\label{eq:X_C}
   X_{C,M}=\sum_{\{i\}}\frac{(N_i-N_{i,iso})(N_{i,M}-N_{i,iso})}{N_{i,iso}}.
\end{align*}
Here $\{i\}$ is a set of angular bins, $N_i$ is the number of events detected in bin $i$, and
$N_{i,M}$ is the average number of events expected to be detected in the $M$ (e.g. isotropic,
unbiased, biased) model. In order to avoid sensitivity to the possible distortion of the UHECR
intensity map by magnetic fields, we used $6^\circ\times6^\circ$ angular bins and excluded the
Galactic plane region, $|b|<12^\circ$. The value of $X_{C,UB}$ can be straight forwardly calculated
using the numerical representations of the UHECR maps at
\url{http://www.weizmann.ac.il/~waxman/criso}.

The $X_C$ statistics is more sensitive to expected anisotropy signature than the commonly used
power spectrum, $C_\ell = \frac{1}{2\ell+1}\sum_{m=-\ell}^{m=\ell} a_{\ell m}^2$
(e.g.~\cite{Sigl:2004yk}), and the two-point correlation function,
$W(D)=\sum_i^N\sum_{j<i}\Theta(D-D_{ij})$ (e.g.~\cite{DeMarco:2006a}). This can be seen from table
\ref{table:compProc}(a). The anisotropy signal is stronger at lower energy: Although the contrast
of the fluctuations in the UHECR intensity is higher at high energy, the signal becomes weaker at
higher energies since the number of observed UHECR drops rapidly with energy, see table
\ref{table:compProc}(b). In \cite{Kashti:2008bw} we also show that a few fold increase of the Auger
exposure is likely to increase the significance to $>99\%$ CL, but not to $>99.9\%$ CL (unless the
UHECR source density is comparable or larger than that of galaxies).

\begin{table}[tb]\vbox{
\begin{tabular}{|c|ccc|}\hline
$E>40$~EeV, 100 events & P(I/UB) & P(I/B) & P(UB/B) \\
  \hline
$X_{C,UB}$ & 23\% & 79\% & 42\% \\
$X_W(\{D=10,20,30,40\})$ & 7\% & 12\% & 10\% \\
$X_C(\{\ell=2\})$ & 6\%& 8\% &7\%\\
\hline
\end{tabular}
\begin{tabular}{|c|ccc|}\hline
 $X_{C,UB}$ & P(I/UB) & P(I/B) & P(UB/B) \\%& P$_{\mbox{Rep}}$\\
  \hline
$E>20{\rm\  EeV}$ (1205 events)& 45\%  &  94\%  &  45\% \\% & $>$99.9 \\
$E>40{\rm\  EeV}$ 300 events  &   39\%  &  94\%  & 52\%\\% & $>$99.9 \\
$E>60{\rm\  EeV}$ (94 events) &  31\%  &  87\%  &  42\% \\% & 97  \\
$E>80{\rm\  EeV}$ (31 events) &  22\%  &  63\%  &  24\% \\% & 65 \\
\hline
\end{tabular}}
\caption { Probabilities $P(M_1|M_2)$ to rule out model $M_1$ at a certain confidence  level (CL),
assuming that the UHECR source distribution follows model $M_2$. Calculated from $10000$ Monte
Carlo realizations for Auger exposure. (a) Comparing different statistics, $X_{C,UB}$,
$X_C({\ell})=\sum_{\ell}(C_\ell-C_{iso,\ell})^2/\sigma_\ell^2$, where $\sigma_\ell^2$ is the
variance, and $X_W$ (defined similarly for the two-point correlation function). (b) Comparing
different energy thresholds with $X_{C,UB}$ statistics.}\label{table:compProc}
\end{table}

Auger reported in \cite{AugerSc:2007} a correlation between the arrival direction distribution of
27 UHECRs of $>5.7\times10^{19}$~eV and between the angular distribution of low-luminosity AGNs
included in the V-C AGN catalog \cite{V-C}. However, the V-C AGN catalog is merely a compilation of
AGN data available in the literature, and is therefore incomplete both in its sky coverage and in
its luminosity coverage,
 %"The V-C catalog should not be used for any statistical
%analysis as it is not complete in any sense, except that it is, we hope, a complete survey of the
%literature" (introduction of ref.~\cite{V-C})
hence the results of \cite{AugerSc:2007} are unclear. Using the $X_{C,UB}$ statistics, we analyzed
\cite{AugerSc:2007} data, see fig.~\ref{fig:Spec+Auger}(b). According to our analysis, the data is
inconsistent with isotropy at $\sim98\%$ CL, and consistent with a source distribution that traces
galaxy density.
%, with some preference to source distribution which is biased with respect to the
%galaxy distribution.
Note, however, that the optimization of the energy threshold made in \cite{AugerSc:2007} raises the
concern that the significance with which isotropy is ruled out maybe
overestimated. %The anisotropy signature should be detectable also $E>4\times10^{19}$~eV.
 %In order to confirm our detection of a correlation with LSS, one may search
%for the anisotropy signature in the Auger data using a lower energy threshold, $4\times10^{19}$~eV,
%or repeat the analysis with an energy threshold of $5.7\times10^{19}$~eV using a larger data set.

% Koers and Tinyakov - one paragraph.

Recently, \cite{Koers:2008ba} compared the $X_{C,UB}$ statistics to an improved KS statistics. They
showed that for most energy thresholds the $X_{C,UB}$ statistics have higher probability to rule
out isotropy (see their table I). However, they get probabilities that are different then ours for
$X_{C,UB}$. This is probably due to the fact that they use a different galaxy redshift catalog, the
catalog of Kalashev et al. \cite{Kalashev:2007ph}, deduced from 2MASS XSC \cite{Jarrett}. We
believe that this catalog is not suitable for the analysis of CR anisotropy, since it includes
mainly photometric redshifts, which have systematic uncertainty of 30\% that depends on the
luminosity, thus distorting the density field. Note that the future ``2MASS Redshift Survey''
\cite{Jarrett} will improve PSCz, by measuring redshifts of 100,000 galaxies till 85~Mpc.

\end{document}